\documentclass[12pt]{iopart}

\usepackage{graphicx}
\begin{document}

\title[Calculations of giant magnetoresistance in Fe/Cr trilayers] 
{ 
Calculations of giant magnetoresistance in Fe/Cr trilayers using 
layer potentials determined from {\it ab-initio} methods}

\author{M. Pereiro$^{1,2}$\footnote{Author to whom any correspondence should be addressed (fampl@usc.es).}
, D. Baldomir$^{1,2}$, S. V. Man'kovsky$^3$, K. Warda$^4$, J. E. Arias$^2$, L. Wojtczak$^4$, and J. Botana$^{1,2}$ }

\address{$^1$Departamento de F\'{\i}sica Aplicada, Universidade de Santiago de Compostela, 
Santiago de Compostela E-15782, Spain. 
}
\address{$^2$Instituto de Investigaci\'ons Tecnol\'oxicas, Universidade de Santiago de Compostela, Santiago de Compostela E-15782, Spain.
}
\address{$^3$Institute for Metal Physics of the National Academy of Sciences of Ukraine, Kiev, Ukraine.}
\address{$^4$Solid State Physics Department, University of {\L}\'od\'z, ul. Pomorska 149/153, 90-236 {\L}\'od\'z, Poland
}

\begin{abstract}
The {\it ab initio} full-potential linearized augmented plane-wave method
explicitly designed for the slab geometry was employed to 
elucidate the physical origin of
the layer potentials for the trilayers $n$Fe/3Cr/$n$Fe(001), where $n$ is
the number of Fe monolayers. The
thickness of the transition-metal ferromagnet has been ranged from $n=1$ up to
$n=8$ while the spacer thickness was fixed to 3 monolayers. The calculated potentials were  
inserted in the Fuchs-Sondheimer formalism in order to calculate the giant 
magnetoresistance (GMR) ratio. The predicted GMR ratio was compared with the experiment and
the oscillatory behavior of the GMR as a function of the ferromagnetic layer thickness
was discussed in the context of the layer potentials.
The reported results confirm that the interface monolayers 
play a dominant role in the intrinsic GMR.  
\end{abstract}

\maketitle

\section{Introduction}
Advances in ultrathin-film fabrication techniques have made possible, only quite recently, the construction of thin magnetic transition metal layers, 
separated by very thin non-magnetic layers (spacers), forming superlattices 
or sandwiches. The multilayer system Fe/Cr/Fe has played a fundamental role 
because the giant magnetoresistance (GMR) was first discovered on it \cite{baibich}.  
Control of the spacer thickness with great accuracy, keeping constant the width of 
the magnetic layers, has been the main preoccupation of experimentalists for a long time, 
 because this factor is one of the basic conditions for obtaining great 
values of GMR as well as the coupling constant between the magnetic layers, but the 
variation of the magnetic layer thickness and its influence
on GMR had remained without experimental measurements until quite 
recently \cite{bloemen,okuno}.
This fact is understandable because most of these experiments were guided by 
the so far existing models in either the quantum-well \cite{edwards1} (QW) or 
the Ruderman-Kittel-Kasuya-Yosida (RKKY) picture \cite{bruno}. These models
study the oscillatory exchange coupling of the magnetic 
layers, mediated by the electrons in the spacer, as a function of the
spacer thickness and remaining the ferromagnetic layer thickness constant
in most of the studied layered structures. However, in this work we focus 
our attention on the GMR effect produced by the variation
of the magnetic layer thickness and  surprisingly we have found an oscillating magnetic 
behavior of the magnetic layers. The explanation resides more in the
presence of QW states than invoking the RKKY-like models, as we will see below.

Nowadays, it is well known that the interface
between layers plays a dominant role in the GMR and obviously it depends
strongly on the materials of the sample \cite{wiatrowski}. However, the controversy
about whether GMR originates from bulk or interface scattering is still open \cite{zahn}. 
This article will threw some light on this controversy because we calculate
the contribution to GMR coming from the bulk and the interface layers. Moreover, we will
show that the magnetic properties change as a function of the 
ferromagnetic layer thickness (FLT) and the contribution to 
the GMR effect of this system is not negligible as we will prove using the Boltzmann model
where the potentials are calculated from the {\it ab initio} data.
Thus, we obtain the density of states (DOS) and energy
bands for different FLT 
and through them we can calculate the potentials. 
A remarkable fact is that the former potentials are numerically close to those
reported by Hood and Falicov \cite{falicov,falicov1} which were chosen without 
an explicit justification. Thanks to considering the full electronic structure
of these materials, we are able to obtain directly
the magnetoresistance of these Fe/Cr/Fe samples in a good agreement with
the experiment, at least for a certain interval of the ferromagnetic
layer thickness \cite{okuno}.
It should be stressed that our  {\it ab initio} calculations and the interpretation 
through Boltzmann formalism can give us information about  some 
confusion areas where the mesoscopic concepts play a fundamental role such as
in the case of the interface.
For systems of reduced dimensions (in this article corresponds to a number
of iron monolayers (ML) less than four) 
the physical magnitudes such as the magnetic moments, conductivity or even the
spin change drastically their behavior, 
having striking consequences for the magnetoresistance effect like for
example a considerable enhancement of the GMR ratio.

In the present article, we investigate the GMR for the layer potentials of
Fe/Cr/Fe trilayers determined by means of density functional theory (DFT)
calculations as a function of the
thickness of the Fe layer. Our study is also restricted to the case of current 
in plane geometry (CIP). Although the study of GMR versus the spacer thickness is
well described in literature, only a few articles are devoted to the case of a variable
ferromagnetic layer. Nevertheless, we have found that the study of this case is 
more suitable for investigating the intrinsic origen of GMR. In order to show and
explain our results we have organized the article as follows: in Sec.~\ref{sec0}
we present the method of calculation. The Fe/Cr/Fe trilayer system is
described and the physical origin of the layer potentials is determined according to
the DFT calculations. Likewise, a brief description of the effective electron masses
is presented in the next subsection. After that, the transport model is introduced
in such a way which allows us to use the data based on DFT calculations, in 
particular, for the relaxation time and the effective mass. The method is
developed in order to describe the electronic transport of the Fe/Cr/Fe
trilayers in analogy to the evolution of the Co/Cu multilayer conductivity \cite{papani}.
In Sec.~\ref{sec2} the results of our calculations are discussed  
and the predictions for GMR properties are reported. Sec.~\ref{sec3} includes some remarks and a brief 
summary of the article.

Thus, the aim of the present article is to describe GMR properties by means of the data
obtained within DFT method and applied to the Fuchs-Sondheimer formalism extended for 
trilayers.

\section{Method of calculation and Computational details}
\label{sec0}
\subsection{\label{sec1}Self-consistent calculations} 
{\it Ab initio} electronic structure calculations were performed within the framework of density functional theory in the local spin density 
approximation (LSDA). The exchange-correlation potential was used in the 
form of Vosko, Wilk, and Nusair \cite{vosko}. As it was reported in
Ref.~\cite{herper}, the use of LSDA instead of the generalized 
gradient approximation (GGA) in Fe/Cr 
trilayers is adequate because GGA improves only the geometry optimization, while
the energy calculations (in particular, Fermi energy) are more or less the 
same. Since we use a fixed 
geometry and the transport properties depend only on the Fermi energy, it is 
not necessary at all to 
use the GGA. The Kohn-Sham
equations \cite{kohn} were solved using the full-potential linearized 
augmented plane wave (FP-LAPW) method in slab 
geometry \cite{cherepin,weinert}. This method is extremely advantageous for 
computing the electronic structure of magnetic multilayers, because it was 
designed to take into account the slab geometry and the interaction between 
the outermost monolayers of the trilayer system and vacuum. Thus, our method guarantees that
all trilayer systems investigated in this work are two-dimensional translational
invariants. To our best knowledge, this is the first attempt to account 
specifically for the slab geometry of layers in contrast with other methodologies
where the two-dimensional invariance is achieved by inserting vacuum layers to separate
the interaction between the slabs \cite{qian}. The valence states 
were calculated in a scalar-relativistic approximation. A grid of 15 {\bf 
k}-points in an irreducible wedge of the 2D BZ was used during 
self-consistent field (SCF) cycles and, once converged, 
a mesh of 45 {\bf k}-points was considered to evaluate the final energy. Inside the 
muffin-tin spheres, basis functions with angular momentum components 
from 1 up to 8 were included. The charge density and potential within the 
muffin-tin spheres were expanded into the lattice harmonics with angular momentum from 1 
up to 6. More than 60 Augmented Plane Waves per atom were used for the variational 
basis set. 
\subsection{Fe/Cr/Fe trilayers} 
\begin{figure}
\includegraphics[width=8cm,angle=0]{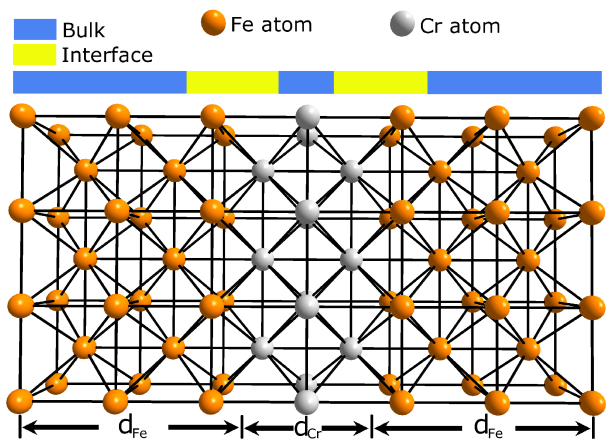}
\caption{\label{fig:1} (Color online) Schematic picture of the generic Fe$_n$/Cr$_3$(001) 
system. Blue and yellow color bar comprise the number of bulk
and interface monolayers, respectively.} 
\end{figure}

Calculations were carried out for the slab consisting of 3 monolayers 
of Cr(001) and a variable number of Fe monolayers (1$\le$$n$$\le$8), where
$n$ is the number of Fe monolayers. The Fe layers are intrinsically ferromagnetic 
and the Cr layers are
intrinsically antiferromagnetic (see column 4 of Tables~\ref{table1} 
and~\ref{table2}). Figure
\ref{fig:1} shows the schematic picture of 3 ML Cr in
between 5 ML Fe for the bcc crystal orientation (001). 
We consider the monolayers to be in x-y plane and stacked along the z direction.
It is well known that the magnetic properties, DOS, and energy bands depend
strongly upon the atomic structures of the thin films. Therefore it is 
necessary to begin with an optimized structure for Fe$_n$/Cr$_3$ system. 
Nevertheless, no 
further attempt was made to relax the lattice parameter, i.e. the lattice 
constant for Fe and Cr was assumed to be Cr 
bcc bulk-like, that is, a$_0$=2.88 \AA~because
we observed small differences in the obtained results
when these constants were taken differently.  

\subsection{Layer potentials}
\label{sec:2c}
The single band model representation leads to the energies
\begin{equation}
	E_{\nu\sigma}(k)=\frac{1}{2}m_\sigma^*(v_\sigma(k))^2+V_{\nu\sigma}
\end{equation}
of electrons moving in the intrinsic potential $V_{\nu\sigma}$ with the spin $\sigma$
in monolayer $\nu$
and velocity $v_{\sigma}(k)$. The intrinsic potentials require for these calculations to be
determined on the basis of {\it ab initio} DFT calculations described in Sec.~\ref{sec1}.
Since these values are the electron energies corresponding to the
bottom of conductivity band, they  were taken to be equal 
to the energies of $s,p$-electrons in the $\Gamma$-point of 2D BZ, since 
we assume that these 
electrons have the main contribution to the conductivity. 
To describe the spin-dependent conductivity within our model, these energies were split by the value $\Delta E^{d-band}$ 
which is equal to the energy 
splitting of majority and minority-spin d-bands in Fe layers while in 
Cr layers the potential values were taken the same for both spin 
directions. Thus, 
\begin{equation} 
\label{eq:eq2} 
V_{\nu \sigma} = E^{s,p-band}_{\nu}({\bf k}=0) - \zeta \sigma \Delta 
  E^{d-band} 
\end{equation} 
where $\zeta=1$ for Fe slab 
and $\zeta=0$ for Cr slab. Note that the spin 
electron in Bohr magneton units is $\sigma=+\frac{1}{2}$ 
for majority electrons and $\sigma=-\frac{1}{2}$ 
for minority electrons 
and the confinement of electrons within two-dimensional 
slab results in quantization of electron states in the 
direction perpendicular to the plane of this slab (z-direction). 
This quantum-size effect is essential, especially when the slab 
thickness is restricted to several atomic layers \cite{ortega,mirbt,perez}. 
Therefore,  we took it into account in our present 
calculations when the 
thickness of Fe film was equal to one-, two- or three atomic 
layers. 
 
\begin{figure}
\includegraphics[width=8cm,angle=0]{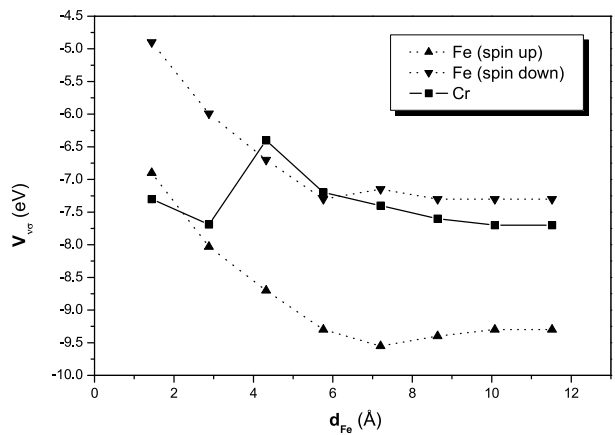}
\caption{\label{fig:2} Dependence of the Fe majority (triangles), Fe minority
(inverted triangles) and Cr (squares) layer potentials
versus the FLT, d$_{\rm Fe}$, for a fixed
Cr thickness, d$_{\rm Cr}=4.32$~\AA. For $n \le 3$, the values of the layer potentials
are given in average.}
\end{figure}

Since the quantum-well electronic states, in the case of ultrathin Fe 
films, are well 
localized in single atomic layer, we assume that the physical properties 
on this layer are determined mainly by the states which have the degree 
of localization in the layer more than $60\%$. The energy of the electron 
state in $\Gamma$-point corresponding to the energy band with high 
localization in the $i$-th atomic layer were taken as the 
$i$-th potential value. In the case of a thicker Fe 
film ($n\ge4$) the electron states are more delocalized and they cannot be related 
to only one atomic layer, therefore we took in these cases 
the values of potentials $V_{\nu \sigma}$ common for a whole Fe 
film. The potential values in Cr film had always the same value
for all Cr atomic layers. The variations of the layer potentials
for $n\ge4$ are expected to be quite small because of the delocalization
of the electron states, such as is shown in Fig.~\ref{fig:2}.

\subsection{Description of the effective mass}

\begin{figure}
\includegraphics{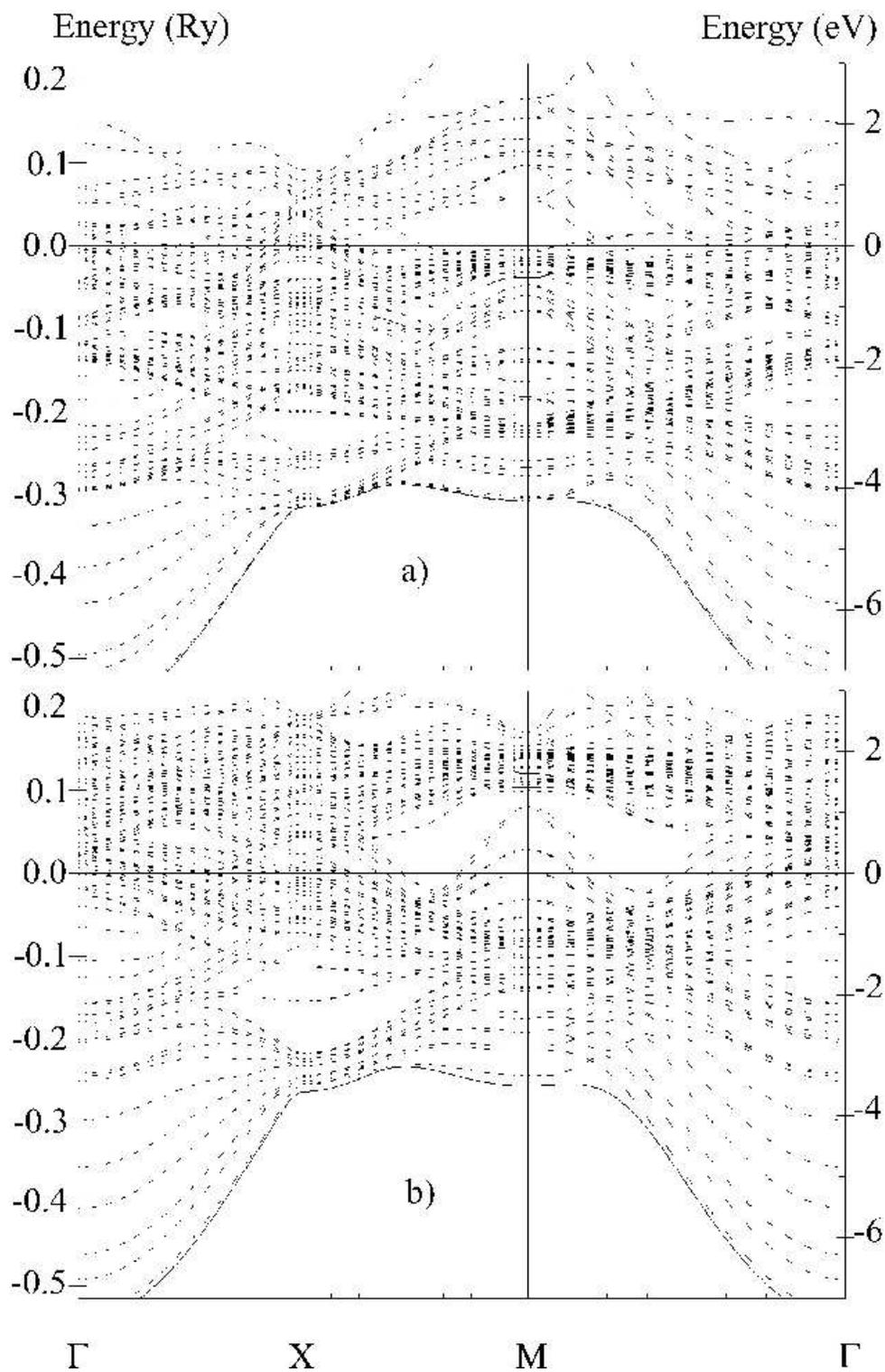}
\caption{\label{fig:3} Energy bands of the majority-spin states (a) and 
the minority-spin states (b) for the system Fe$_5$/Cr$_3$. Solid lines
represent the states with more than 50\%~ of localization in the interface
Fe layer. The Fermi level is at 0 eV.} 
\end{figure}

The effective mass parameter $m^*_\sigma$ can be calculated by means of the
standard procedure of DFT applied to the band structure calculations. An example of the
complicated spin-dependent band structure provided by our DFT calculations is
showed in Fig.~\ref{fig:3} for the system Fe$_5$/Cr$_3$. The effective masses have been 
calculated as the second derivative of the s,p electronic band energy $E_\sigma(k)$ close
to the Fermi level with respect to the Fermi wave vector $k_F$ and evaluated on the
$\Gamma$-point, according to the following relation
\begin{equation}
	\frac{1}{m^*_\sigma}=\frac{1}{\hbar^2}\left(\frac{d^2 E_\sigma(k)}{dk^2}\right)_{k=k_F}.
\end{equation}
The energy band close to the Fermi level could always be approximated by a second-order polynomial 
in all considered systems. This argument is what underpins the parabolic approximation adopted 
in Sec.~\ref{sec:2c}.
The numerical evaluation of the effective masses is collected in Table~\ref{table3}. We
can observe that our calculated effective masses are close to 4$m_e$, that is the value
assumed for the effective mass in Fe and Cr \cite{falicov,visscher}.
To the best of our knowledge, no effective masses has been reported for Fe/Cr multilayers.
In layered materials, we have only found the effective masses for Cu thin films deposited 
on the fcc Co film and they are in good agreement with the ones reported in 
table~\ref{table3} \cite{johnson}. 

\subsection{The Transport Model}  

In order to consider the transport properties using the DFT data, the electronic
transport for the Fe/Cr/Fe trilayers can be described by the
Boltzmann formalism \cite{falicov,falicov1,camley,barnas,warda,ziese} applied to the 
current in plane geometry. In every monolayer, the electric current is then determined
by appropriate distribution functions in terms of the velocity $v_{z}$ in the
direction z perpendicular to the interface for the electron with spin up and spin
down due to the translational symmetry in the plane of the film. The electrons involved
in transport are embedded within the potentials $V_{\nu \sigma}$ of each monolayer $\nu$.
These potentials were determined using the DFT data in conjunction with 
Eq.~(\ref{eq:eq2}). 

With the aim of calculating the conductivity we write the Boltzmann equation 
in the relaxation time $\tau_{\nu\sigma}^*$ approximation for the distribution
functions $g_{\nu\sigma }^{\beta }$ ($\beta=+,-$) which are determined with respect to the
electric field applied in the direction of the electric current.
The solutions are found taking into account the boundary conditions at the outer surfaces, namely
$g_{1 \sigma}^{+}=P_{1\sigma} g _{1 \sigma}^{-}$ (at z=0) and 
$g_{n \sigma}^{-}=P_{n\sigma} g_{n \sigma}^{+}$ (at z=$2(n+1)a_0$)
with $g_{\nu\sigma}^+$ and $g_{\nu\sigma}^-$ standing for the solutions with $v^{z}\geq 0$ and 
$v^{z}<0$, respectively.
In order to extend our considerations to a more realistic case from the physical
point of view, we define the specularity factor $P_{\nu\sigma}$ on the basis of the physical
picture of the electrons involved in transport crossing through the potential barrier. In this
case the value of $P_{\nu\sigma}$ corresponds to the specularity factor in the 
Fuchs-Sondheimer \cite{fuchs,sondheimer} conductivity theory of thin films and
leads to the effective result
\begin{equation}
\label{eq:eq3}
P_{\nu\sigma}(\theta)=\frac{1-\frac{(\chi_{\nu\sigma})^{2}\cos ^{4}\theta +2}
{1+(\chi_{\nu\sigma})^{2}\cos ^{4} \theta}}{1-\frac{(\chi_{\nu\sigma})^{2}
\cos^{4}\theta +2}{(\chi_{\nu\sigma})^{2}\cos ^{4} \theta}\exp 
\left( - \frac{a_0}{\tau^*_{\nu\sigma} v^{2}\cos \theta} \right)} 
\end{equation}
for $\nu$ ranging from 1 up to $2n+3$, that is an alternative version to other 
models which assume the coefficients
for coherent transmission and specular reflections determined quantum mechanically
by matching free electron wave functions and their derivatives at each 
interface (see Fig.~\ref{fig:1}). In Eq.~(\ref{eq:eq3}),  
$\theta$ is the angle of incidence of electrons measured with respect to the
z-axis. The parameter $\chi_{F}^{\sigma}$ is defined in terms of the Fermi 
velocity $v_{F}$ as 
\begin{equation}
   \chi_{\nu\sigma}=\frac{2 m^*_{\sigma} \tau^*_{\nu\sigma} v _{F}^{2}}{\hbar} 
\end{equation}
where $m_{\sigma}^*$ is the effective mass of $\sigma$ electrons. From
the physical point of view $\chi_{\nu\sigma}$ represents the
ratio of the electron free path with respect to the de Broglie wavelength.

The distribution functions $g_{\nu\sigma}^\alpha$ depend on the
relaxation times $\tau_{\nu\sigma}^*$ which can be evaluated in terms of
the DFT calculations by means of the Fermi golden rule. The relaxation
time can be expressed as follows \cite{kubler}:
\begin{equation}
	\label{eq:rel_time}
	(\tau_{\nu\sigma}^*)^{-1}=\frac{c}{\hbar}  \rho_\nu(\varepsilon_F)(V_{\nu\sigma})^2
\end{equation}
where $\rho_\nu(\varepsilon_F)$ is the local density of states per monolayer $\nu$ at the Fermi energy, and
c is the number of the scattering centers relative to the total number of atoms
and it plays the role of a
calibration factor in order to compare it with the average values in the
case of samples discussed in literature. The  Eq.~(\ref{eq:rel_time})
is important for thin films due to the presentation of the relaxation time
distribution across a sample.

The total current in CIP geometry along the direction $\alpha$ defined by the 
electric field $\mathbf{E}^\alpha$ 
is obtained after averaging the current density
over the whole thickness of the film and it is given by the relation \cite{falicov,falicov1}
\begin{equation} 
J^{\alpha}=\frac{-|e|}{d}  
\sum_{\nu=1,\sigma=\uparrow\downarrow}^{2n+3} 
\left \lbrack \frac{m_{\sigma }^*}{2\pi\hbar}\right \rbrack^3
\int_0^d  
\int v^{\alpha }g_{\nu \sigma}^\beta (\mathbf{v},z,E^{\alpha })d\mathbf{v} dz 
\end{equation} 
where $e$ is the electric charge and $d$ is the length of the sample
in the z direction. Assuming that 
the total conductivity of a sample can be defined as  $\Sigma=1/\rho=
\left(dJ^{\alpha }/dE\right)_{E=0}$, and the magnetoresistance (MR) ratio as
$\Delta \rho/\rho_s=\left(\rho _{\uparrow\downarrow }-\rho _{\uparrow \uparrow })\right/
\rho _{\uparrow\downarrow }$, then the theoretical predictions of the GMR can 
be obtained straightforwardly. It is worth to emphasize that the MR ratio is
found by calculating independently the resistivities for the parallel
($\rho_{\uparrow \uparrow }$) and the antiparallel ($\rho _{\uparrow \downarrow}$)
alignment of the magnetic moments in adjacent magnetic layers. The present approach
remains in analogy to the calculations of Zahn {\it et al.} \cite{papani} who use
the semiclassical Boltzmann theory with a spin independent relaxation time 
approximation. However, in our case, the relaxation time is not only a spin 
dependent but also shows a local character.

Concerning the problem of whether GMR originates from bulk or interface scattering,
we consider the contribution to MR from bulk and interface layers separately. An example to
allocate the bulk and interface layers
where bulk and interface scattering occurs is plotted
in Fig.~\ref{fig:1}. We have discretized the conductivity layer by layer
with appropriate boundary conditions. Thus, the summation of the conductivity of the bulk layers
and the conductivity of the interface layers was done independently to obtain the bulk 
and interface contributions to MR.
\section{Results and discussion}
\label{sec2}
\subsection{Layer magnetic moments and potentials}
\label{sec2:a}
First, we describe briefly the computational results obtained using
the LAPW method. The numerical results of our calculations along with the
layer potentials calculated according to Eq.~(\ref{eq:eq2}) and the
relaxation times provided by Eq.~(\ref{eq:rel_time}) are presented in
Table~\ref{table1} and Table~\ref{table2}.\footnote{For the sake of
  notation, a
brief comment on Table~\ref{table1} and Table~\ref{table2} is required: Cr(1)
stands for the monolayer of Cr that is closest to a Fe monolayer; Cr(2) is
the
central monolayer of the Cr slab, that is, the second of the three Cr
monolayers; Fe(1) stands for the external monolayer of the Fe Film; and
Fe(2),
Fe(3), ..., are the corresponding monolayers inside the Fe slab.} 

\begin{figure} \includegraphics[width=8cm,angle=0]{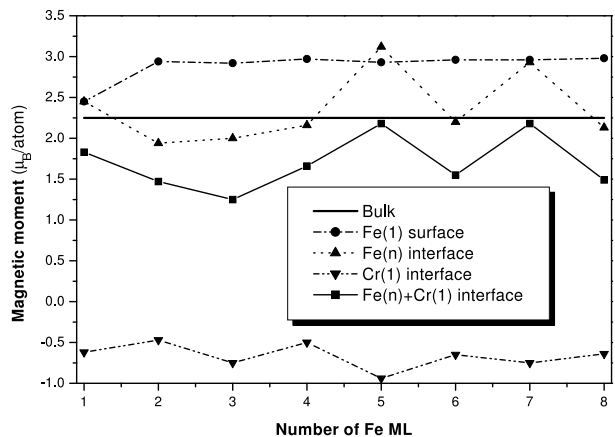} 
\caption{\label{fig:4} Behavior of the surface (circles) and interface
(triangles) Fe magnetic moment versus Fe film thickness. Inverted triangles
represent the value of the Cr interface magnetic moment and the squares 
correspond to the numerical sum of the Fe and Cr interface magnetic moments. Solid
line is the magnetic moment of the Fe bulk which is equal to 2.25 $\mu_B$.}
\end{figure}

According to Fig.~\ref{fig:4}, the
behavior of the magnetic moment versus the Fe film thickness 
for the interface Fe layer is non-monotonic
in contrast with the surface Fe magnetic
moment which ranges monotonically from 2.45 $\mu_B$ for $n=1$ up to 2.98 $\mu_B$ for
$n=8$  (see fourth column of Table~\ref{table1} and
Table~\ref{table2} for Fe(1) ML).
We have also found the existence of highly localized
states at Fe/Cr interface when the Fe film
thickness is not too small ($n \geq 4$).
These states can result in
the onset of a big magnetic moment in the interface Fe layer as it
takes place also at the surface.
However, the localization of these states near the interface means
that they should not be affected by the thickness of Fe film and the magnetic
moments near Fe/Cr interface should be independent of Fe film
thickness. This is a feature that in our opinion,
is due to the QW states, which are delocalized throughout
the slab and their energy
changes together with increase of the Fe film thickness \cite{mankovsky}.
The QW states have a resonance with the interface states for
$n=4,6,8$. Consequently, the interface
states become less localized which results in a decrease of their
exchange splitting as well as in a decrease of the interface Fe magnetic moment. The
opposite behavior is found for the case $n=5,7$ leading to an increase
of the interface Fe magnetic moment, such as is shown in Fig.~\ref{fig:4}. 

\begin{figure}
\includegraphics[width=8cm,angle=0]{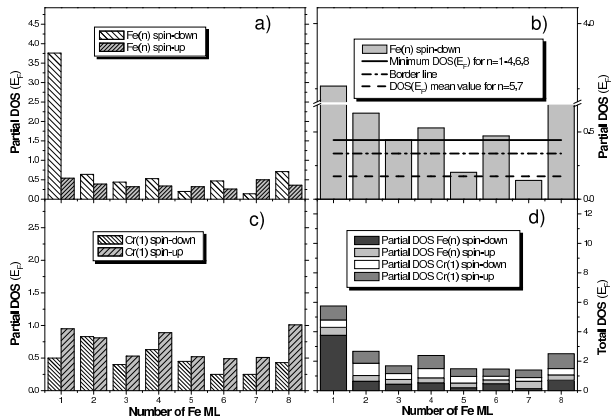}
\caption{\label{fig:5} Column graph exhibiting the partial DOS($\varepsilon_F$) versus
the Fe layer thickness for (a) the Fe interface  ML spin-up and down, 
(b) Fe interface ML spin-down and (c) Cr interface ML spin-up and down 
polarization. (d) show the total DOS($\varepsilon_F$) as function of the FLT.
The dash dot line in (b) establishes a reasonable criterion for 
considering the minority spin channels insignificant for the case $n=5,7$.} 
\end{figure}

In this paragraph, our computational results are put in relation with the GMR effect, 
with the aim of enable deductions about the effects that enhance the GMR, at least from a 
qualitatively point of view.
The quantitative predictions
will be showed in Sec.~\ref{sec:3b}. For $n=5,7$ we have a special behavior on
the Fe/Cr interface (see Table~\ref{table2}), which deserves a more detailed comment.
The first characteristic that we obtain, different 
from what happens for $n=1,2,3,4,6,8$,
is that we observe that the majority spins are the main contribution at the Fermi level
(see Fig.~\ref{fig:5}(a)). 
Notice that the DOS($\varepsilon_F$) minority spin for $n=1,2,3,4,6,8$ has the corresponding
values 3.76, 0.64, 0.44, 0.53, 0.47, and 0.71, while for $n=5,7$ 
they are 0.21 and 0.14, respectively. As can be inferred, these values are much 
lower than the previous ones (in Fig.~\ref{fig:5}(b), we 
can see clearly
the criterion employed to consider the magnetic channels to be insignificant
for $n=5,7$, where the dash dot line is the border line between 
negligible and appreciable channels).
In fact, the lowest value for n=1,2,3,4,6 and 8 is 0.44, which is above the dash
dot line. Thus, we observe that 
the minority spin
channels are very small for these two special cases in comparison with the ones seen
above.
The above comments lead us to conclude that
our electronic structure calculations within DOS column 
graphs presented in Fig.~\ref{fig:5}
show that there is 
essentially only one magnetic channel (semimetallic) on the 
iron-chromium interface for $n=5,7$ (see 
Fig.\ref{fig:5}(a)-(b) and Table~\ref{table2}) and a very high increase
of the magnetism on its interfaces, since the other channel is
the one of the minority spin density, quasiequal to zero. Nevertheless, for $n=1,2,3,4,6$, and 8 
we obtain both kinds of magnetic polarizations: up and down (see 
Table~\ref{table1}). This situation is very interesting for the 
GMR effect if we accept that this phenomenon is mainly due to the scattering of the
electrons on these interfaces (a feature which will be discussed
in Sec.~\ref{sec:3b}), that is, the spin-polarized electrons on the one hand,
enhance their velocity when they find the magnetic polarizability of the impurity
parallel to the magnetic layer, but on the other hand, the spin-polarized electrons
diminish their velocity when they find the polarizability in opposite direction.
Therefore, we have a spin valve effect in which there is a selection of one 
of the two possible magnetic channels, and this drives 
a decrease or an increase
of the conductivity, depending on whether the electrons are parallel or
antiparallel to the magnetic polarization of the interface.

\subsection{GMR predictions}
\label{sec:3b}
\begin{figure}
\includegraphics[width=8cm]{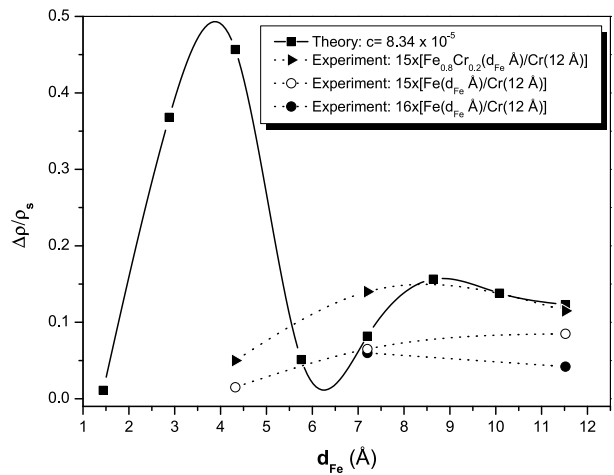}
\caption{\label{fig:6} Plot of the calculated (solid line)
and experimental \cite{okuno} (dashed lines) MR ratio as a function of the 
ferromagnetic layer thickness.} 
\end{figure}

So far, the models based on the Boltzmann equation specially developed for describing
the multilayer transport properties introduce the potentials as a parameter or 
consider them in a semiempirical way \cite{falicov,falicov1}. The value of 
this potential is usually taken as a constant however for ultrathin films 
the potential depends on the thickness of the sample as it was 
reported in Ref.~\cite{warda1} and consequently, the 
electronic transport is influenced by the thickness of the 
sample (not only the thickness of the spacer but also the thickness of 
the ferromagnetic layer). Contrary to the usual procedure described in literature, 
in this article, the values of the potentials as a function of the FLT 
were determined using the DFT methods described earlier (see 
Fig.~\ref{fig:2}, Table~\ref{table1} and Table~\ref{table2}). The 
dependence of the MR ratio versus the thickness of the ferromagnetic 
layer d$_{\rm Fe}$ is shown in Fig.~\ref{fig:6}.
The values of the MR ratio  obtained for the thickness dependent potential 
are indicated by full rectangles and they were calculated for a dilute
concentration of scattering centers, that is, for c=$8.34 \times 10^{-5}$ \cite{kubler}.
The calculated MR exhibits an oscillatory
behavior for a thickness in the range of the x axis displayed by Fig.~\ref{fig:6},
although the oscillations appear at wrong periodicity compared with
the experiments reported in Ref.~\cite{okuno}. In the light of the present calculations,
we can conclude that the oscillatory character of the GMR comes mainly from a variable potential
distribution through out the thin film, since that the oscillatory behavior of the GMR is
hidden as long as the potential is allowed to be constant, like for example in the case of
the semiclassical model in which every layer is represented by a quantum well with a constant
potential \cite{falicov}. The MR ratio shown in Fig.~\ref{fig:6} exhibits two maxima: the first
one at around 4~\AA~and the second one close to the 8~\AA. Thus, although the period
of oscillations is wrong compared to the experiment \cite{okuno}, the second maximum agrees quite well with
the measurements and we can observe in Fig.~\ref{fig:6} that the theoretical and experimental
values remain in good agreement in the interval of d$_{\rm Fe}\ge 8$ \AA. Only the proper 
treatment and choice of the potential allows us to obtain the oscillatory 
behavior even in the semiclassical approach \cite{pereiro1}. The important problem, which 
is up till now omitted in the literature, is the determination of the 
potential at the interface or even proper treatment of the interface. It is 
a well known fact that at the interface there exists a significant change 
of the potential but this change is not of the abrupt character. At the 
interface there is a transition region where the interdiffusion and the spin 
mixing take place while the change of the potential should be continuous 
according to the background. 

We note that even though
on one hand the GMR effect depends strongly on the temperature \cite{gijs} and on another hand
we are comparing our data (0 K) with the ones reported by Okuno {\it et al.} (77 K) at different
temperatures, the influence of the temperature in the interval [0,77] can be considered
negligible \cite{gijs}. The discrepancy of our results
with the experimental ones is less than about 10-15~\% for $n \ge 4$, which represents an
improvement with respect to other {\it ab-initio} calculations \cite{schep}. The aforementioned
small discrepancy can be due to the high number of layers (multilayers) 
considered in the experiment
in comparison with the reduced number of layers (trilayer) of our Fe/Cr sample. For
the sake of comparison with the results reported by Okuno {\it et al.}, it is 
noteworthy that a remarkable
message from our results is the low-dependence behavior of GMR with the number of
Fe/Cr layers because the average value of the GMR ratio in function of the Fe
layer thickness is around 0.1 for $n\ge 4$ in our
calculations and in the experimental results. In order to explain this low-dependence
behavior, our proposal consists in considering a quasi-linear dependence of 
resistance of different number of Fe/Cr layers. Thus, after calculating the GMR ratio, 
the proportional constant is divided by itself having no appreciable effect in GMR.
This tendency disappears for the case of two or three iron monolayers
leading to a considerable GMR ratio as shown in Fig.~\ref{fig:6}, since that
the electron states are more localized and thus, they favor an increase of the 
layer potentials.

\begin{figure}
\includegraphics[width=8cm]{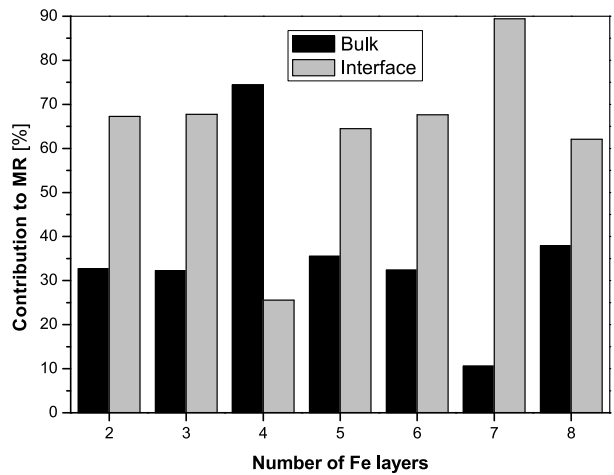}
\caption{\label{fig:7} Bulk and interface contribution to the total MR ratio versus
the thickness of the ferromagnetic layer for c=$8.34 \times 10^{-5}$.}
\end{figure}

We should remind the reader  
that studies of transport in metallic 
superlattices (and particularly for trilayers) are affected by many inherent complexities of the material. 
Many possible complications arise in these types of artificial material, 
among them, interfacial interdiffusion at various lateral
length scales, \cite{fullerton,colino,velez} bulk defects, structural changes as a function 
of an individual layer and overall thicknesses, different length scales 
affecting the structure, the boundary conditions on the outermost layers and differences in the 
magnetotransport along the different directions in the superlattices. But 
the key point in the mechanism of GMR is the relative importance of bulk and 
interfacial scattering. Measurements as a function of layer thickness have 
claimed that the GMR originates from the bulk and that the interfacial 
roughness does not play a crucial role \cite{mattson,qiu} even in the 
current perpendicular to plane configuration. Other measurements, 
in which the interface was modified by the addition of small amounts of 
interfacial impurities, claim that the interfacial scattering plays a 
dominant role \cite{parkin}. 
The bulk and interface contribution to the 
total MR ratio as a function of the Fe monolayers is showed in Fig.~\ref{fig:7}.
The results, except for the case n=4, clearly show that the monolayers that belong 
to the interface dominate the GMR because the potential abruptly changes in this 
region and it becomes in an important source of scattering. However, in the case n=4 
since the potential
barrier is  less important than the other cases (see in Fig.~\ref{fig:2} the square and
the inverted-triangle for n=4), the 
influence of the bulk monolayers become more important. Likewise, in this anomalous case,
the reduced contribution 
of the interface scattering favors a diminishment of the total MR, that is reflected in 
Fig.~\ref{fig:6}.

The last measurements reported in Ref.~\cite{santamaria}
and extensive comparative studies of
the growth, structure, magnetization and magnetotransport in Fe/Cr 
superlattices show that the intrinsic GMR originates from interfacial scattering and 
is determined by the interface width \cite{pereiro1}. This experimental fact confirms 
the results shown in Fig.~\ref{fig:7} and our 
earlier conclusion that the interface should be treated physically not as an 
ideal plane but as a transition zone between different materials where we 
can observe a mixture of two compounds (interdiffusion) and where not only a 
significant change of the potential but also a different kind of magnetism 
take place \cite{wiatrowski}. The DFT results \cite{pereiro} 
discussed in Sec.~\ref{sec2:a} confirm the important role of interface, specially 
the significant increase of magnetic moment at interface 
for $n=5,7$ (see Fig.~\ref{fig:4}). This 
fact finds its reflection in a significant change 
of the potential at interface for different spin polarizations 
(Fig.~\ref{fig:2} and Fig.~\ref{fig:5}). The difference in interface potential 
constitutes the main physical reason of different scattering of 
electrons with different 
spin orientation and it is the  main mechanism responsible for GMR in multilayers. 

\section{Conclusion}
\label{sec3}  
We have shown that using the two combined methods: the {\it ab initio} 
methodology for the accurate
calculation of the potentials and the semiclassical approach based on
the Fuchs-Sondheimer formalism,  
we obtain  the GMR values for the trilayer Fe/Cr/Fe system which 
oscillates versus the thickness of the ferromagnetic 
layers. In literature, the GMR oscillations versus the
nonmagnetic spacer are emphasized while 
the role of  ferromagnetic layer is considered sporadically. In ordinary
semiclassical approach 
the oscillations of this kind do not exist, however our calculations
can predict this oscillatory behavior for
ultrathin layers. Our results based on
DFT calculations for Fe/Cr system emphasize the very 
important role of the FLT
via thickness dependent potential as well as the contribution of the
interface to the GMR ratio. It is worthwhile to notice that the
present approach does not contain any semiempirical parameter 
apart from the calibration constant c.

\ack
The authors thank the supercomputer center CESGA
for access to supercomputers. One of
us (M. P.) acknowledges partial support by the Xunta de Galicia, under 
the project No. PGIDIT02TMT20601PR.

\newpage
\begin{table}  
\caption{\label{table1}Computational results for Fe$_n$/Cr$_3$ trilayers 
(n=1-4,6,8) along with the layer potentials calculated according to Eq.~(\ref{eq:eq2})
and the relaxation times provided by Eq.~(\ref{eq:rel_time}) with c=$8.34\times 10^{-5}$. 
The symbol m denotes magnetic moment per atom, $\rho_\nu(\varepsilon_F)$ is 
the density of states at the 
Fermi level, and V$_{\nu\sigma}$ represents the potential of the majority and 
minority spin per monolayer.}
\footnotesize\rm
\begin{tabular*}{\textwidth}{@{}l*{15}{@{\extracolsep{0pt plus12pt}}l}}
\br
Trilayer &E$_F$&ML& m & \multicolumn{2}{c}{$\rho_\nu(\varepsilon_F)$} & \multicolumn{2}{c}{V$_{\nu\sigma}$}
& \multicolumn{2}{c}{$\tau_{\nu\sigma}^*$}\\
&(eV) & & $\left(\frac{\mu_B}{\mbox{at.}}\right)$    & 
\multicolumn{2}{c}{$\left(\mbox{eV}^{-1}\right)$} & 
\multicolumn{2}{c}{(eV)} & \multicolumn{2}{c}{$\left(10^{-13} \mbox{s}\right)$}\\
& & & & $\uparrow$ & $\downarrow$ & $\uparrow$ & $\downarrow$ & $\uparrow$ & $\downarrow$\\
\mr
&&Fe(1)  &  2.45 & 0.54 & 3.76 & -6.90 & -4.90 & 3.07 & 0.87\\
Fe$_1$/Cr$_3$&-4.32& Cr(1) & -0.62 & 0.95 & 0.50 & -7.30 & -7.30 & 2.35 & 1.58 \\
             & &Cr(2) &  0.62 & 0.63 & 0.94 & -7.30 & -7.30 & 1.56 & 2.96\\
\hline
&&Fe(1)&2.94 & 0.13 &2.88 & -8.19 & -5.83 & 9.05 & 0.81\\
&&Fe(2)&1.94 & 0.39 &0.64 & -7.86 & -6.14 & 3.28 & 3.27 \\
Fe$_2$/Cr$_3$ &-3.97 &Cr(1) &-0.47& 0.81 &0.83 & -7.69 & -7.69 & 1.65 & 1.61  \\
&&Cr(2)&0.35& 0.33 &0.50 & -7.69 & -7.69 & 4.04 & 2.67 \\
\hline
&&Fe(1)&2.92 & 0.14 & 2.12 & -8.7 & -6.3 & 7.45 & 0.94 \\
&&Fe(2)&2.43 & 0.15 & 0.62 & -8.5 & -6.4 & 7.28 & 3.11 \\
Fe$_3$/Cr$_3$&-4.32&Fe(3) &2.00 & 0.32 & 0.44 & -8.3 & -6.7 & 3.58 & 4.00 \\
&&Cr(1)&-0.75& 0.53 & 0.40 & -6.3 & -6.3 & 3.75 & 4.97\\
&&Cr(2)& 0.71& 0.21 & 0.23 & -6.3 & -6.3 & 4.47 & 8.65\\
\hline
&&Fe(1)&2.97 & 0.17 & 2.32 & -9.3 & -7.3 & 5.37 & 0.45\\
&&Fe(2)&2.27 & 0.28 & 0.34 & -9.3 & -7.3 & 3.26 & 4.36\\
&&Fe(3)&2.49 & 0.26 & 0.57 & -9.3 & -7.3 & 3.51 & 2.60\\
Fe$_4$/Cr$_3$&-4.32&Fe(4)&2.16 & 0.34 & 0.53 & -9.3 & -7.3 & 2.68 & 2.79 \\
&&Cr(1)&-0.50& 0.89 & 0.63 & -7.2 & -7.2 & 1.71 & 2.42\\
&&Cr(2)&0.50 & 0.25 & 0.40 & -7.2 & -7.2 & 6.09 & 3.81\\
\hline
&&Fe(1)&2.96 & 0.08 & 1.72 & -9.4 & -7.3 & 11.16 & 0.86\\
&&Fe(2)&2.25 & 0.23 & 0.36 & -9.4 & -7.3 & 3.88 & 4.11\\
&&Fe(3)&2.41 & 0.38 & 0.28 & -9.4 & -7.3 & 2.35 & 5.29\\
&&Fe(4)&2.25 & 0.41 & 0.34 & -9.4 & -7.3 & 2.18 & 4.36\\
Fe$_6$/Cr$_3$&-3.95&Fe(5)&2.34 & 0.28 & 0.21 & -9.4 & -7.3 & 3.19 & 7.05 \\
&&Fe(6)&2.20 & 0.26 & 0.47 & -9.4 & -7.3 & 3.44 & 3.15\\
&&Cr(1)&-0.65& 0.49 & 0.25 & -7.6 & -7.6 & 2.79 & 5.47\\
&&Cr(2)&0.58 & 0.52 & 0.48 & -7.6 & -7.6 & 2.63 & 2.85\\
\hline
&&Fe(1)&2.98 & 0.04 & 1.91 & -9.3 & -7.3 & 22.81 & 0.78\\
&&Fe(2)&2.19 & 0.14 & 0.36 & -9.3 & -7.3 & 6.52 & 4.11\\
&&Fe(3)&2.35 & 0.10 & 0.44 & -9.3 & -7.3 & 9.13 & 3.37\\
&&Fe(4)&2.19 & 0.14 & 0.34 & -9.3 & -7.3 & 6.52 & 4.36\\
&&Fe(5)&2.26 & 0.13 & 0.30 & -9.3 & -7.3 & 7.02 & 4.94\\
Fe$_8$/Cr$_3$&-3.97&Fe(6)&2.25 & 0.17 & 0.39 & -9.3 & -7.3 & 5.37 & 3.80 \\
&&Fe(7)&2.35 & 0.17 & 0.32 & -9.3 & -7.3 & 5.37 & 4.63\\
&&Fe(8)&2.13 & 0.36 & 0.71 & -9.3 & -7.3 & 2.54 & 2.09\\
&&Cr(1)&-0.64& 1.01 & 0.43 & -7.7 & -7.7 & 1.32 & 3.10\\
&&Cr(2)&0.49 & 0.29 & 0.46 & -7.7 & -7.7 & 4.59 & 2.89\\
\br
\end{tabular*}
\end{table}
\begin{table}  
\caption{\label{table2}Computational results for Fe$_n$/Cr$_3$ trilayers
(n=5, 7) along with the layer potentials calculated according to Eq.~(\ref{eq:eq2})
and the relaxation times provided by Eq.~(\ref{eq:rel_time}) with c=$8.34\times 10^{-5}$.
The symbols represent the same than in Table~\ref{table1}.}
\footnotesize\rm
\begin{tabular*}{\textwidth}{@{}l*{15}{@{\extracolsep{0pt plus12pt}}l}}
\br
Trilayer &E$_F$&ML& m & \multicolumn{2}{c}{$\rho_\nu(\varepsilon_F)$} & \multicolumn{2}{c}{V$_{\nu\sigma}$}
& \multicolumn{2}{c}{$\tau_{\nu\sigma}^*$}\\
&(eV) & & $\left(\frac{\mu_B}{\mbox{at.}}\right)$    & 
\multicolumn{2}{c}{$\left(\mbox{eV}^{-1}\right)$} 
& \multicolumn{2}{c}{(eV)} & \multicolumn{2}{c}{$\left(10^{-13} \mbox{s}\right)$}\\
& & & & $\uparrow$ & $\downarrow$ & $\uparrow$ & $\downarrow$ & $\uparrow$ & $\downarrow$ \\
\mr
&&Fe(1)&2.93 & 0.17 & 2.50 & -9.4 & -7.3 & 5.04 & 0.61\\
&&Fe(2)&2.35 & 0.27 & 0.92 & -9.4 & -7.3 & 3.31 & 1.61\\
&&Fe(3)&2.41 & 0.39 & 0.29 & -9.4 & -7.3 & 2.29 & 5.11\\
Fe$_5$/Cr$_3$&-4.56&Fe(4)&1.79 & 0.48 & 0.79 & -9.4 & -7.3 & 1.86 & 1.88\\
&&Fe(5)&3.12 & 0.32 & 0.21 & -9.4 & -7.3 & 2.79 & 7.05\\
&&Cr(1)&-0.94& 0.52 & 0.45 & -7.4 & -7.4 & 2.77 & 3.20\\
&&Cr(2)&0.64 & 0.20 & 0.14 & -7.4 & -7.4 & 7.21 & 10.30\\
\hline 
&&Fe(1)&2.96 & 0.17 & 2.70 & -9.3 & -7.3 & 5.37 & 0.55\\
&&Fe(2)&2.24 & 0.46 & 0.44 & -9.3 & -7.3 & 1.98 & 3.37\\
&&Fe(3)&2.29 & 0.47 & 0.22 & -9.3 & -7.3 & 1.94 & 6.73\\
&&Fe(4)&2.21 & 0.52 & 0.37 & -9.3 & -7.3 & 1.76 & 4.00\\
Fe$_7$/Cr$_3$&-4.02&Fe(5)&2.27 & 0.62 & 0.13 & -9.3 & -7.3 & 1.47 & 11.39 \\
&&Fe(6)&1.77 & 0.55 & 0.32 & -9.3 & -7.3 & 1.66 & 4.63\\
&&Fe(7)&2.93 & 0.50 & 0.14 & -9.3 & -7.3 & 1.83 & 10.58\\
&&Cr(1)&-0.75& 0.51 & 0.25 & -7.7 & -7.7 & 2.61 & 5.32\\
&&Cr(2)&0.50 & 0.49 & 0.46 & -7.7 & -7.7 & 2.72 & 2.89\\
\br
\end{tabular*}  
\end{table}
\begin{table}  
\caption{\label{table3}The spin-dependent effective mass parameters in units of the free
electron mass ($m_e$).} 

\begin{indented}
\lineup
\item[]\begin{tabular}{@{}*{3}{l}}
\br
Trilayer & $m^*_\uparrow$ & $m^*_\downarrow$\\
&(m$_e$) & (m$_e$)\\
\mr
Fe$_1$/Cr$_3$&3.15&3.98\\
Fe$_2$/Cr$_3$&3.55&4.59\\
Fe$_3$/Cr$_3$&3.56&4.62\\
Fe$_4$/Cr$_3$&3.85&5.08\\
Fe$_5$/Cr$_3$&3.92&5.13\\
Fe$_6$/Cr$_3$&3.95&5.17\\
Fe$_7$/Cr$_3$&3.93&5.20\\
Fe$_8$/Cr$_3$&3.93&5.20\\
\br
\end{tabular}
\end{indented}
\end{table}

\end{document}